\documentclass[twocolumn,twoside,slac]{revtex4}
\usepackage{graphicx}
\usepackage{fancyhdr}
\begin{document}

\title{The probability of making a correct decision in hypotheses testing
as estimator of quality of planned experiments\\

\bigskip

\hfill  S.I. Bityukov, N.V. Krasnikov}


\begin{abstract}
{In the report the approach to estimation of quality of planned 
experiments is considered. This approach is based on the analysis of 
uncertainty, which will take place under the future hypotheses testing 
about the existence of a new phenomenon in Nature. 
The probability of making a correct decision in hypotheses testing is 
proposed as estimator of quality of planned experiments. 
This estimator allows to take into account systematics and statistical
uncertainties in determination of signal and background rates.}

\end{abstract}


\maketitle

\thispagestyle{fancy}

\section{Introduction}

One of the common goals in the forthcoming experiments is the search for new 
phenomena. In estimation of the discovery potential of the planned experiments
the background cross section (for example, the Standard Model cross section)
is calculated and, for the given integrated luminosity $L$, the average 
number of background events is $n_b = \sigma_b \cdot L$.
Suppose the existence of new physics 
leads to additional nonzero signal cross section $\sigma_s$ with the same 
signature as for the background cross section  that 
results in  the prediction of the additional average number of signal events
$n_s = \sigma_s \cdot L$ for the integrated luminosity $L$.
The total average number of the events is 
$<n> = n_s + n_b = (\sigma_s + \sigma_b) \cdot L$.
So, as a result of new physics existence, we expect an excess 
of the average number of events. Let us suppose the probability of the 
realization 
of $n$ events in the experiment is described by function $f(n;\lambda)$
with parameter $\lambda$. 

In the report the approach to estimation of quality of planned 
experiments is considered. This approach is
based on the analysis of uncertainty, which will take 
place under the future hypotheses testing about the existence of
a new phenomenon in Nature. 

We consider a statistical hypothesis 
\begin{center}
$H_0$:  {\it new physics is present in Nature}
\end{center}
against an alternative hypothesis 
\begin{center}
$H_1$: {\it new physics is absent in Nature}. 
\end{center}
The value of uncertainty is defined by the values of the probability to 
reject the hypothesis $H_0$ when it is true  (Type I error)
\begin{center}
$\alpha$ = $P(reject~H_0|H_0~is~true)$ 
\end{center}
\noindent
and the probability to accept the hypothesis $H_0$ when the hypothesis 
$H_1$ is true  (Type II error)
\begin{center}
$\beta$ = $P(accept~H_0|H_0~is~false)$. 
\end{center}
\noindent
Here $\alpha$ is a significance of the test and $1-\beta$ is a power
of the test. 

We propose to use as estimator of the quality of planned experiments
the probability of making a correct decision in the future hypotheses testing 
$1 - \hat \kappa$

\begin{equation}
\displaystyle 1 - \hat \kappa = 1 - \frac{\hat \alpha + \hat \beta}{2}, 
\end{equation}

\noindent
and as estimator of the distinguishability of the hypotheses 
$1 - \tilde \kappa$

\begin{equation}
\displaystyle 1 - \tilde \kappa = 
1-\frac{\hat \alpha + \hat \beta}{2-(\hat \alpha + \hat \beta)}, 
\end{equation}

\noindent
where $\hat \alpha$ and $\hat \beta$ are the estimators of 
Type I error $\alpha$ and Type II error $\beta$ calculated by the applying
of the equal-tailed test ($\hat \alpha = \hat \beta$). 

\noindent
The $1 - \hat \kappa$ is the estimator of quality of planned experiments. 
This estimator allows to take into account systematics and statistical
uncertainties~\cite{b2} in determination of signal and background rates. 
The $1 - \hat \kappa$ have no dependence on the choice what is $H_0$, 
what is $H_1$. This value is free from restrictions of such type. 
It is an advantage of our approach. We also propose to use 
an equal probability test~\cite{b1} as a good approximation 
of the equal-tailed test for estimation
of the probabilities $\hat \alpha$ and $\hat \beta$ in the case of
Poisson distributions. 

\section{What is meant by the probability of making a correct decision in 
hypotheses testing?}

Suppose that the probability of the realization of $n$ events in 
experiment is described by the function $f(n;\lambda)$ with parameter 
$\lambda$ and we know the expected number of signal events $n_s$ and expected 
number of background events $n_b$.

Let us determine what we mean by the probability of making a correct 
decision under the future hypotheses testing about the presence or
absence of the new phenomenon in Nature in case of carrying out the
planned experiment. Let us use the frequentist approach, i.e. consider
all the possible results of the experiment in cases when both the
hypothesis $H_0$ is true or the hypothesis $H_1$ is true, define the
criterion for the hypothesis choice and calculate the probability
of making a correct decision. It is possible, because we construct the
critical area in such a way that the probability of incorrect
and, correspondingly, correct choice
in favour of one of the hypothesis have no dependence on whether true
is $H_0$ or $H_1$. 
So, we will consider 2
conditional distributions of probabilities~(see, Fig.1)

\begin{equation}
\cases {f_0(n) = f(n; n_s+n_b), \cr
        f_1(n)=f(n; n_b)} 
\end{equation}

\noindent
making numerical calculations. We suppose that any 
prior suppositions about $H_0$ and $H_1$ can be included to
$f_0(n)$ and $f_1(n)$.

\begin{figure}
 \begin{center}
            \resizebox{8.cm}{6.5cm}{\includegraphics{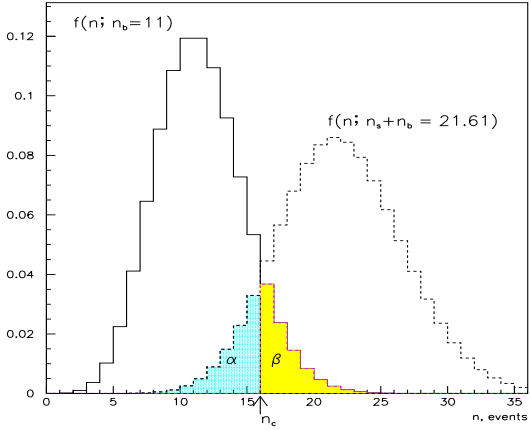}} 
\caption{Equal probability test for the case $n_b=11$ and $n_s=10.61$ 
gives the critical value $n_c=16$ and, correspondingly, the probability of 
uncorrect decision $\hat \kappa = 0.09$ and the measure of distinquishability
of hypotheses $\tilde \kappa = 0.1$.}
    \label{fig:1} 
 \end{center}
\end{figure}

After choosing a critical value
(or a critical area) same way, it is possible to count up the 
estimators of Type I error
$(\hat \alpha)$ and Type II error $(\hat \beta)$. 

In the case of applying the equal-tailed test their combination
\begin{equation}
\displaystyle \hat \kappa = \frac{\hat \alpha + \hat \beta}{2} 
\end{equation}

\noindent 
is the probability of making incorrect choice in favour of one of the
hypothesis.

The explanation is very simple. In actuality we must estimate the random
value $\kappa = \alpha + \beta = \hat \kappa + e$, where $\hat \kappa$ is
a constant term and $e$ is a stochastic term.
The $\alpha$ is a fraction of incorrect 
decisions if the hypothesis $H_0$ takes place. In this case the $\beta$
is absent because the hypothesis $H_1$ is not realised in Nature.
Correspondingly, the $\beta$ is a fraction of incorrect decisions if the 
hypothesis $H_1$ takes place. In this case the $\alpha$ is absent. 
Let us the hypothesis $H_0$ be true then the Type I error equals 
$\hat \alpha$ and the error of our estimator (Eq.4)
is $\hat e = \hat \kappa - \hat \alpha = 
\displaystyle \frac{\hat \alpha + \hat \beta}{2} - \hat \alpha  =
-\frac{\hat \alpha - \hat \beta}{2}$. If hypothesis $H_1$
is true then the Type II error equals 
$\hat \beta$ and the error of the estimator 
is $\hat e = \hat \kappa - \hat \beta = 
\displaystyle \frac{\hat \alpha + \hat \beta}{2} - \hat \beta  =
\frac{\hat \alpha - \hat \beta}{2}$. By this mean the stochastic term 
takes the values $\displaystyle \pm\frac{\hat \alpha - \hat \beta}{2}$ 
and if we require $\hat \alpha = \hat \beta$
then both errors of the estimation are equal to 0~~~
($\hat \kappa - \hat \alpha = \hat \kappa - \hat \beta = 0$). As a result
the estimator (Eq.4) gives the probability of making an
incorrect decision in future hypotheses testing.
Really, if $\hat \alpha = \hat \beta$ and the $H_0$ takes place in Nature then
$\hat \kappa = (\hat \alpha + \hat \beta)/2 = 2 \cdot \hat \alpha / 2 =
\alpha$. In the same manner $\hat \kappa = \beta$ if the $H_1$ takes place.
Accordingly, $\displaystyle 1 - \hat \kappa$ is the probability 
to make a correct choice under the given critical value.

Under the hypotheses testing we can also estimate the measure
$1 - \tilde \kappa$ of distinguishability 
of the hypotheses $H_0$ and $H_1$~\footnote{If 
we will use the geometric approach 
(let us the $A$ is a set of possible realizations
of the result of the planned experiment if the hypothesis $H_0$ takes place
in Nature and the $B$ is a set of possible realizations of the result of the 
planned experiment if the hypothesis $H_1$ takes place)
then we have the total number of the possibilities for decision equals 
to $A \bigcup B$ and the fraction of incorrect decisions will be 
$\displaystyle \tilde \kappa = \frac{A \bigcap B}{A \bigcup B} =
\frac{\hat \alpha + \hat \beta}{2-(\hat \alpha + \hat \beta)}$. 
} by the calculation of

\begin{equation}
\displaystyle \tilde \kappa = 
\frac{\hat \alpha + \hat \beta}{2 - (\hat \alpha + \hat \beta)}. 
\end{equation}

\noindent
There are 3 possibilities.

\begin{itemize}
\item Distributions $f_0(n)$ and $f_1(n)$ have no overlapping, hence,
the distributions are completely distinguishable and any result of the
experiment will give the correct choice between hypotheses,
i.e. $\tilde \kappa = 0$.

\item Distributions $f_0(n)$ and $f_1(n)$ coincide completely. 
It means, that it is impossible to get a correct answer, i.e.
$f_0(n)$ and $f_1(n)$ are not distinguishable, i.e. $\tilde \kappa = 1$.

\item Distributions $f_0(n)$ and $f_1(n)$ do not coincide, but
they have an overlapping, i.e. $\tilde \kappa$ is ratio of the
probability of making incorrect choice to probability
making correct choice in favour of one of the hypothesis. 

\end{itemize}

\section{The choice of critical area}

Let the probability of the realization of $n$ events in the experiment 
be described by Poisson distribution with parameter $\lambda$, i.e.

\begin{equation}
f(n; \lambda)  = \frac{{\lambda}^n}{n!} e^{-\lambda}.
\end{equation}

In this case the estimators of Type I and II errors, which will take place 
in testing of hypothesis $H_0$ versus hypothesis $H_1$, can be written as 
follows:

\begin{equation}
\cases{\displaystyle \hat
\alpha = \sum^{n_c}_{i = 0}{f(i; n_s + n_b)} = \sum^{n_c}_{i = 0}{f_0(i)}, \cr
\displaystyle \hat
\beta = 1 - \sum^{n_c}_{i=0}{f(i; n_b)} = 1 - \sum^{n_c}_{i=0}{f_1(i)},}
\end{equation}

\noindent
where $n_c$ is a critical value. 
Correspondingly, the magnitude 
$\displaystyle \hat \kappa = \frac{\hat \alpha + \hat \beta}{2}$
will have minimal value under applying of the equal probability 
test~\cite{b1} with critical value~(see, Fig.1)

\begin{equation}
n_c = \displaystyle [\frac{n_s}{ln(n_s+n_b) - ln(n_b)}],
\end{equation}

\noindent
where square brackets mean the integer part of a number.
It is easy to show that the $\hat \kappa$ has a minimum if we require
$f_0(i) = f_1(i)$ (for discrete distributions it corresponds to condition 
$f_0(i) \le f_1(i)$), i.e.
\begin{equation}
\displaystyle \frac{n_b^{n_c}e^{-n_b}}{n_c!} =
\displaystyle \frac{(n_s+n_b)^{n_c}e^{-(n_s+n_b)}}{n_c!}. 
\end{equation}

\noindent
It is direct consequence of the equation

\begin{equation}
\displaystyle \hat \kappa = \frac{\hat \alpha + \hat \beta}{2} = 
\frac{1}{2}(1 - \sum^{n_c}_{i = 0}{(f_1(i) - f_0(i))}).
\end{equation}

\noindent
The value of $\hat \kappa$ decreases with increasing of $i$ from $0$ 
up to $i=n_c$. As soon as $f_0(i) > f_1(i)$ the value of $\hat \kappa$ 
increases.

Note that the equal probability test gives the results close to the 
results of the equal-tailed test in the case of Poisson distributions 
and we use it hereafter. 
 
Following the given discourse, we can choose
critical values so that $\hat \kappa$ could be minimal and the
probability of correct decision $\displaystyle 1 - \hat \kappa$ - 
maximum for any pair of distributions. As a result it is possible to say,
that the value $\displaystyle 1 - \hat \kappa$ under the optimum 
choice of critical value characterises the quality of planned experiment.

Notice, that such approach works for arbitrary 
distributions~(see, Fig.2), including multidimentional ones.

\begin{figure}[htbp]
  \resizebox{12pc}{!}{\includegraphics{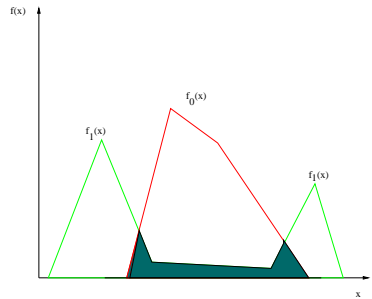}}
\caption{The estimation of uncertainty in hypotheses testing
for arbitrary distributions.}
\end{figure}

\section{How to take into account the statistical uncertainty in the 
determination of $n_s$ and $n_b$?}

Let the values $n_s = \hat n_s$ and $n_b = \hat n_b$ be known from Monte
Carlo calculations. In this case they are random variables. These values
can be considered as estimators of unknown parameters. 
Consequently, the values $n_c$, $\alpha$ and $\beta$ are also
random variables. It means that $1 - \hat \kappa$ is the 
estimator of the probability of making a correct decision 
in hypotheses testing. Let us consider how the uncertainties in
the knowledge of $n_s$ and $n_b$ influence the value of magnitude of the 
Probability of Making a Correct Decision in hypotheses testing (PMCD) 
$1 - \hat \kappa$. Suppose, as before, that the streams of signal and 
background events are Poisson's.

Let us write down the density of Gamma distribution $\Gamma_{a, n + 1}$ 
as~\footnote{Here the traditional designations of Gamma-distribution 
$\displaystyle \frac{1}{\beta}$, $\alpha$ and $x$ is replaced by 
$a$, $n+1$ and $\lambda$, correspondingly.}

\begin{equation}
g_n(a,\lambda) = \displaystyle 
\frac{a^{n+1}}{\Gamma(n+1)} e^{-a\lambda} \lambda^{n},   
\end{equation}

\noindent
where  $a$ is a scale parameter, $n + 1 > 0$ is a shape parameter, 
$\lambda > 0$ is a random variable, and $\Gamma(n+1) = n~!$ 
is a Gamma function. 

Let us set $a = 1$, then for each $n$ a continuous function

\begin{equation}
g_n(\lambda) = \displaystyle \frac{\lambda^n}{n!} e^{-\lambda},~ 
\lambda > 0,~n > -1  
\end{equation}

\noindent
is the density of Gamma distribution $\Gamma_{1, n + 1}$
with the scale parameter  $a = 1$ (see Fig.3). 
The mean, mode, and variance of 
this distribution are given by  $n+1,~n$, and $n+1$, respectively.

As it follows from the article~\cite{Jaynes} (see, also~\cite{Frodesen})
and is clearly seen from the identity~\cite{b4} 
(Fig.4)

\begin{equation}
\displaystyle
\sum_{n = \hat n + 1}^{\infty}{f(n; \lambda_1)} +
\int_{\lambda_1}^{\lambda_2}{g_{\hat n}(\lambda) d\lambda} + 
\sum_{n = 0}^{\hat n}{f(n;\lambda_2)} = 1~,~~~i.e.
\end{equation}

\begin{center}
$\displaystyle
\sum_{n = \hat n + 1}^{\infty}{\frac{\lambda_1^ne^{-\lambda_1}}{n!}} +
\int_{\lambda_1}^{\lambda_2}
{\frac{\lambda^{\hat n}e^{-\lambda}}{\hat n!}d\lambda}
+ \sum_{n = 0}^{\hat n}{\frac{\lambda_2^ne^{-\lambda_2}}{n!}} = 1~$ 
\end{center}

\noindent
for any $\lambda_1 \ge 0$ and 
$\lambda_2 \ge 0$,
the probability of true value of parameter of Poisson distribution
to be equal to the value of $\lambda$ in the case of one
observation $\hat n$ has probability density of 
Gamma distribution $\Gamma_{1,1+\hat n}$. The Eq.(13)
shows that we can mix Bayesian and frequentist probabilities in the given
approach. This identity does not leave a place for any prior except 
uniform. The bounds $\lambda_1$ and $\lambda_2$ fix it.

\noindent

\begin{figure}[htbp]
\caption{The behaviour of the probability density of the true
value of parameter $\lambda$ for the Poisson distribution 
in case of $n$ observed events versus  $\lambda$ and  $n$. Here 
$f(n;\lambda)=g_n(\lambda)=\displaystyle \frac{\lambda^n}{n!}e^{-\lambda}$ 
is both the Poisson distribution with the parameter $\lambda$ 
along the axis  $n$ and the Gamma distribution with a shape 
parameter $n+1$ and a scale parameter 1 along the axis $\lambda$.}
\includegraphics[height=.4\textheight]{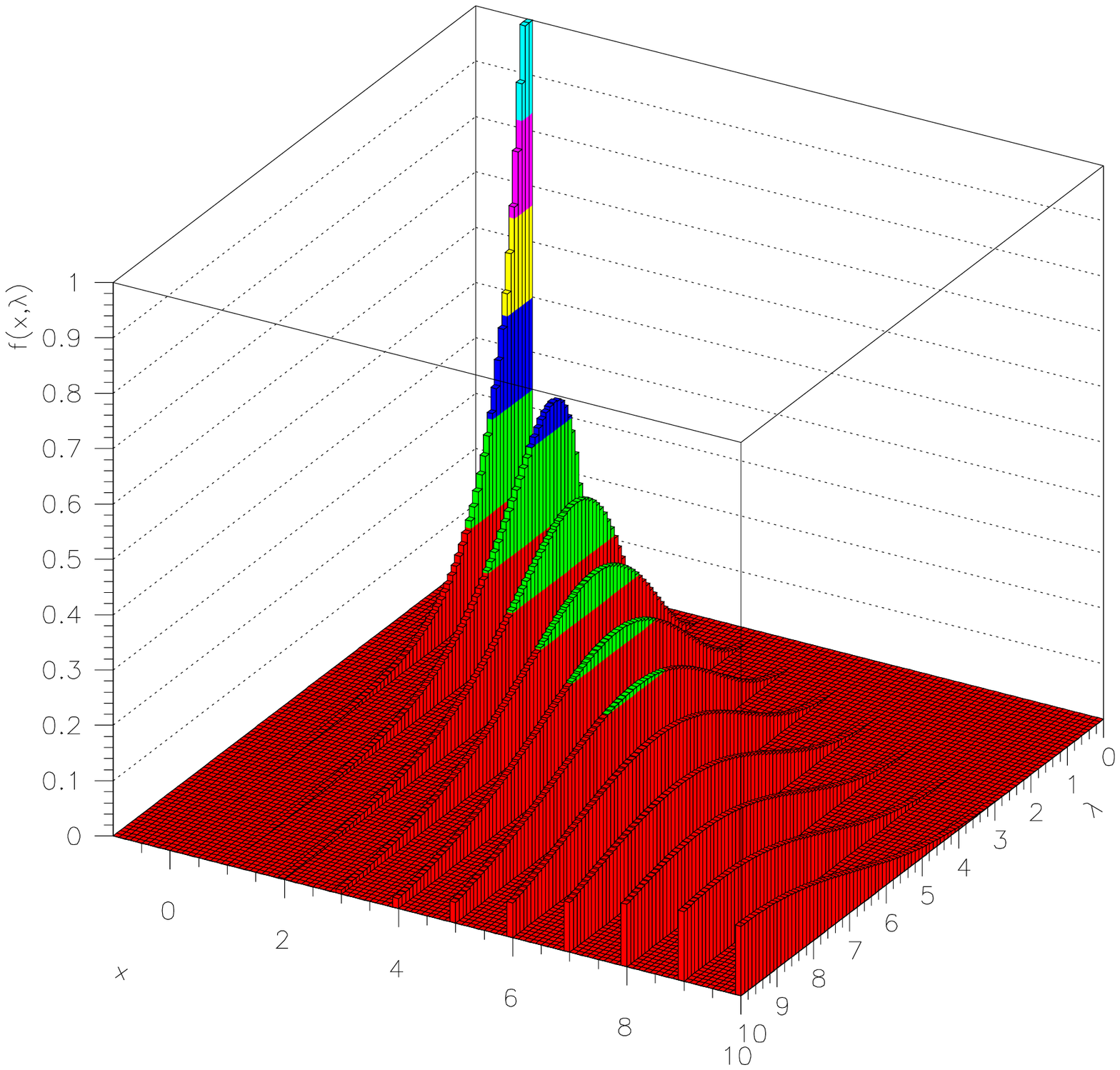}
\end{figure}

\begin{figure}[htbp]
\caption{The Poisson distributions $f(n,\lambda)$
for $\lambda$'s determined by the confidence limits 
$\hat \lambda_1 = 1.51$ and  $\hat \lambda_2 = 8.36$ 
in case of the observed number of events $\hat n = 4$ 
are shown. The probability density of Gamma distribution 
with a scale parameter  $a=1$ and a shape parameter  
$n+1=\hat n+1=5$ is shown within this confidence interval.}
\includegraphics[height=.4\textheight]{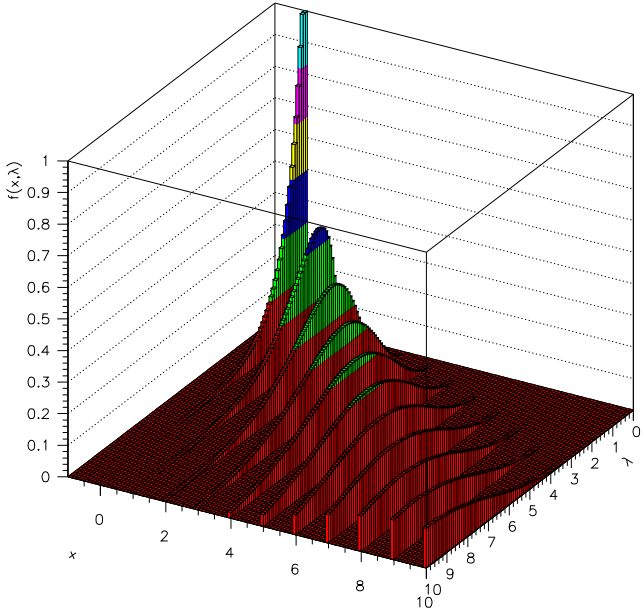}
\end{figure}
                
It allows to transform the probability distributions 
$f(i; n_s+n_b)$ and $f(i; n_b)$ accordingly
to calculate the probability of making a correct decision

\begin{equation}
\cases{\displaystyle \hat \alpha = 
\int_{0}^{\infty}{g_{n_s+n_b}(\lambda)\sum_{i=0}^{n_c}f(i;\lambda)d\lambda} 
= \sum_{i=0}^{n_c} \frac{C^i_{n_s+n_b+i}}{2^{n_s+n_b+i+1}}, \cr
\displaystyle \hat \beta = 1 -
\int_{0}^{\infty}{g_{n_b}(\lambda)\sum_{i=0}^{n_c}f(i; \lambda)d\lambda}
= 1 - \sum_{i=0}^{n_c} \frac{C^i_{n_b+i}}{2^{n_b+i+1}}, \cr 
\displaystyle 1 - \hat \kappa = 1 - \frac{\hat \alpha + \hat \beta}{2}.}
\end{equation}

\noindent
Here the critical value $n_c$ under the future 
hypotheses testing about the observability is chosen in accordance with
test of equal probability (Eq.8) and 
 $C^i_N$ is $\displaystyle \frac{N!}{i!(N-i)!}$.
Also we suppose that the Monte Carlo luminosity   
is exactly the same as the data luminosity later in the experiment.

The Poisson distributed random values 
have a property: if $\xi_i \sim Pois(\lambda_i),~i=1,2, \dots, m$ then 
$\displaystyle \sum^m_{i=1} \xi_i \sim 
\displaystyle Pois(\sum^m_{i=1} \lambda_i)$. 
It means that if we have $m$ observations 
$\hat n_1$, $\hat n_2$, $\dots$, $\hat n_m$ of the same random value 
$\xi \sim Pois(\lambda)$, 
we can consider these observations as one observation 
$\displaystyle \sum^m_{i=1} \hat n_i$  of the Poisson distributed random value 
with parameter $m \cdot \lambda$. According to Eq.(13)
the probability of true value of parameter of this
Poisson distribution has probability density of 
Gamma distribution $\displaystyle \Gamma_{1,1+\sum_{i = 1}^{m}{\hat n_i}}$.
Using the scale parameter $m$ one can show that the probability of true value 
of parameter of Poisson distribution in the case of $m$ observations
of the random value $\xi \sim Pois(\lambda)$ 
has probability density of Gamma distribution 
$\displaystyle \Gamma_{m,1+\sum_{i = 1}^{m}{\hat n_i}}$, i.e. 
(see Eq.(11))

$G(\sum{\hat n_i},m,\lambda) = 
g_{(\sum_{i = 1}^{m}{\hat n_i})}(m,\lambda) = $
\begin{equation} 
\displaystyle \frac{m^{(1+\sum_{i = 1}^{m}{\hat n_i})}}
{(\sum_{i = 1}^{m}{\hat n_i})!} e^{-m\lambda} 
\lambda^{(\sum_{i = 1}^{m}{\hat n_i})}. 
\end{equation}

Let us assume that the integrated luminosity of planned experiment is $L$
and the integrated luminosity of Monte Carlo data is $m \cdot L$. 
For instance, we can divide the Monte Carlo data into $m$ parts with
luminosity corresponding to the planned experiment. The result of Monte Carlo 
experiment in this case looks as set of $m$ pairs of numbers 
$(~(n_b)_i,~(n_b)_i+(n_s)_i~)$, where $(n_b)_i$ and $(n_s)_i$ are the numbers 
of background and signal events observed in each part of Monte Carlo data. 
Let us denote 
$\displaystyle N_b = \sum_{i=1}^m{(n_b)_i}$ and 
$\displaystyle N_{s+b} = \sum_{i=1}^m{((n_s)_i+(n_b)_i)}$.
Correspondingly~(see page 98,~\cite{Frodesen}),

\begin{equation}
\cases{
\displaystyle \hat \alpha = 
\int_{0}^{\infty}{G(N_{b+s},m,\lambda)\sum_{i=0}^{n_c}
f(i;\lambda)d\lambda} = \cr
\displaystyle
\sum_{i=0}^{n_c} C^i_{N_{s+b}+i} \frac{m^{1+N_{s+b}}}{(m+1)^{1+N_{s+b}+i}},\cr 
\displaystyle \hat \beta = 1 -
\int_{0}^{\infty}{G(N_b,m,\lambda)\sum_{i=0}^{n_c}f(i; \lambda)d\lambda} = \cr
\displaystyle
1 - \sum_{i=0}^{n_c} C^i_{N_b+i} \frac{m^{1+N_b}}{(m+1)^{1+N_b+i}}.}
\end{equation}
 
\noindent
As a result, we have a generalized system of equations for the case
of different luminosity in planned data and Monte Carlo data to 
calculate the PMCD
$\displaystyle 1 - \hat \kappa = 1 - \frac{\hat \alpha + \hat \beta}{2}$.
The set of values 
$\displaystyle C^i_{N+i} \frac{m^{1+N}}{(m+1)^{N+i+1}},~i=0,1,\dots$ 
is a negative binomial (Pascal) distribution with real parameters 
$N+1$ and $\displaystyle \frac{1}{m+1}$, 
mean value $\displaystyle \frac{1+N}{m}$ and 
variance $\displaystyle \frac{(1+m)(1+N)}{m^2}$.

\section{A possible way to take into account the systematics}

We consider here forthcoming experiments to search for new physics. 
In this case we must take into account the systematic uncertainty 
which have theoretical origin without any statistical 
properties. For example, two loop corrections for most reactions at present 
are not known. It means that we can only estimate the scale of 
influence of background uncertainty on the observability of signal, i.e. we 
can point the admissible level of uncertainty in theoretical calculations
for given experiment proposal.

Suppose uncertainty in the calculation of exact 
background cross section is determined by parameter $\delta$, i.e. the exact 
cross section lies in the interval $(\sigma_b, \sigma_b (1+\delta))$ 
and the exact value of average number of background events 
lies in the interval 
$(n_b, n_b (1+\delta))$. Let us suppose $n_b \gg n_s$. In this instance
the discovery potential is the most sensitive to the systematic uncertainties.
As we know nothing about possible values of average number of 
background events, we consider the worst case~\cite{b3}. 
Taking into account Eqs.(7)  we have the formulae~\footnote{Eqs.(17) 
realize the worst case when the background 
cross section $ \sigma_b(1 +\delta)$ 
 is the maximal one, but we think that both the signal and the background 
cross sections are minimal. Also, we suppose that 
$n_b(1+\delta) < n_s+n_b$.}

\begin{equation}
\cases{\hat \alpha = \displaystyle \sum^{n_c}_{i = 0}{f(i; n_b + n_s)}  \cr
\hat \beta =  1 - \displaystyle \sum ^{n_c}_{i=0} f(i; n_b(1+\delta)) \cr
1 - \hat \kappa = 1 - \displaystyle \frac{\hat \alpha + \hat \beta}{2},
}
\end{equation} 

\noindent
where $n_c$ is

\begin{equation}
n_c = 
\displaystyle [\frac{n_s - n_b\cdot\delta}{ln(n_s+n_b) - ln(n_b\cdot(1+\delta))}].
\end{equation}

\section{Conclusions}

In this paper we have considered the probability of making a correct 
decision in hypotheses testing to estimate the quality of planned experiments.
This estimator allows to measure the distinguishability  of models.  
We estimate the influence of statistical uncertainty in determination 
of mean numbers of signal and background events and propose a possible
way to take into account effects of one-sided systematic errors.

\begin{acknowledgments}

{The authors are grateful to V.A.~Matveev and V.F.~Obraztsov for 
the interest and useful comments, S.S.~Bityukov, 
Yu.P.~Gouz, G.~Kahrimanis,  A.~Nikitenko, V.V.~Smirnova, V.A.~Taperechkina
for fruitful discussions and E.A.~Medvedeva for help in preparing
the paper. The authors wish to thank the referee of JHEP. This work has 
been supported by grant RFBR 03-02-16933. }

\end{acknowledgments}

\bigskip

\noindent
Sergei I. Bityukov, Division of experimental physics,
Institute for high energy physics, 142281 Protvino, Russia\\
E-mail: {\tt Serguei.Bitioukov@cern.ch, bityukov@mx.ihep.su} \\

\noindent
Nikolai V. Krasnikov, Department of high energy physics,
Institute for nuclear research RAS, 
Prospect 60-letiya Octyabrya 7a, 117312 Moscow, Russia\\
E-mail: {\tt Nikolai.Krasnikov@cern.ch}


\begin{thebibliography}{99}

\bibitem{b1} S.I.Bityukov and N.V.Krasnikov,
{\it On the observability of a signal above background},
Nucl.Instr.\&Meth. {\bf A452} (2000) 518.

\bibitem{b2} S.I.Bityukov, 
{\it On the Signal Significance in the Presence of 
     Systematic and Statistical Uncertainties}, JHEP 09 (2002) 060,~~
     http://www.iop.org/EJ/abstract/1126-6708/2002/09/060;~~ 
e-Print: hep-ph/0207130. 

\bibitem{b3} S.I.~Bityukov and N.V.~Krasnikov,
{\it New physics discovery potential in future experiments,}
Modern Physics Letters {\bf A13} (1998) 3235. 

\bibitem{Jaynes} E.T.~Jaynes: Papers on probability, statistics and
statistical physics, Ed. by R.D. Rosenkrantz, D.Reidel Publishing Company,
Dordrecht, Holland, 1983, p.165.

\bibitem{Frodesen} A.G.Frodesen, O.Skjeggestad, H.Toft, 
{\it Probability and Statistics in Particle Physics,}
UNIVERSITETSFORLAGET, Bergen-Oslo-Tromso, 1979, p.408.

\bibitem{b4}
S.I. Bityukov, N.V. Krasnikov, V.A. Taperechkina,
{\it Confidence intervals for Poisson distribution parameter,}
Preprint IFVE 2000-61, Protvino, 2000; also,
e-Print: hep-ex/0108020, 2001. 

\end{thebibliography}
\end{document}